\documentclass[11pt]{amsart}
\usepackage{geometry}                
\usepackage{graphicx}
\usepackage{amssymb}
\usepackage{epstopdf}
\usepackage{hyperref}
\usepackage{changes}     
\usepackage[square,numbers]{natbib}
\title[Parameter estimation for second order diffusion processes]{Parameter estimation for partially observed second-order diffusion processes}
\author{Jan Albrecht \and Sebastian Reich}
\address{Institut f\"ur Mathematik, Universit\"at Potsdam, Karl-Liebknecht-Str. 24/25, 14476 Potsdam}
\address{Institut f\"ur Physik und Astronomie, Universit\"at Potsdam, Karl-Liebknecht-Str. 24/25, 14476 Potsdam}
\date{\today}                                           

\begin{document}

\begin{abstract}
Estimating parameters of a diffusion process given continuous-time observations of the process via maximum likelihood approaches or, online, via stochastic gradient descent or Kalman filter formulations constitutes a well-established research area. It has also been established previously that these techniques are, in general, not robust to perturbations in the data in the form of temporal correlations of the driving noise. While the subject is relatively well understood and appropriate modifications have been suggested in the context of multi-scale diffusion processes and their reduced model equations, we consider here an alternative but related setting where a diffusion process in positions and velocities is only observed via its positions. In this note, we propose a simple modification to standard stochastic gradient descent and Kalman filter formulations, which eliminates the arising systematic estimation biases. The modification can be extended to standard maximum likelihood approaches and avoids computation of previously proposed correction terms.
\end{abstract}

\maketitle

%
\section{Introduction}
%
We consider the problem of online learning parameters $\theta \in \mathbb{R}^{d_\theta}$ 
of second-order stochastic differential equations (SDEs) 
\begin{subequations} \label{eq:SDE}
\begin{align}
{\rm d}U_t &= f(X_t,U_t){\rm d}t + F(X_t,U_t)\theta {\rm d}t + \sqrt{\sigma}{\rm d}W_t,\\
{\rm d}X_t &= U_t {\rm d}t
\end{align}
\end{subequations}
from incoming observations $X_t^\dagger \in \mathbb{R}^d$, $t\ge 0$, of an underlying reference process. Here $f:\mathbb{R}^{d}\times \mathbb{R}^d\to \mathbb{R}^d$ and $F:\mathbb{R}^d\times \mathbb{R}^d
\to \mathbb{R}^{d\times d_\theta}$ are given functions, $\sigma >0$ is the diffusion constant, and $W_t$ denotes $d$-dimensional Brownian motion. 
The data $X_t^\dagger$ is being generated from a second-order SDE of the same type, i.e.,
\begin{subequations} \label{eq:data}
\begin{align}
{\rm d}U_t^\dagger &= g(X_t^\dagger,U_t^\dagger){\rm d}t + \sqrt{\sigma}{\rm d}W_t^\dagger,\\
{\rm d}X_t^\dagger &= U_t^\dagger {\rm d}t
\end{align}
\end{subequations}
with the drift $g(x,u)$ being unknown. 

Note that for $f, F, g$ independent of position, the system corresponds to a first-order diffusion process where only the integral of the process is being observed.

We assume that $g(x,u)$ is such that the reference process $(U_t^\dagger, X_t^\dagger)$ is ergodic with invariant distribution $\pi$. The optimal parameter $\theta_\ast$ can then be determined by minimizing the ergodic least squares functional
\begin{equation} \label{eq:least squares}
\mathcal{L}(\theta) = \frac{1}{2}
\mathbb{E}_\pi \left[ \|g(X,U) - f(X,U) - F(X,U)\theta\|^2\right],
\end{equation}
which has a unique minimizer provided the Fisher information matrix $\mathbb{E}_{\pi}[F(X, U)^T F(X,U)]$ has full rank. Moreover, we assume that the unique minimizer satisfies 
\begin{equation}
g(x,u) = f(x,u) + F(x,u)\theta_\ast\,.
\end{equation}
However, the gradient of $\mathcal{L}(\theta)$ cannot be computed since $g(x,u)$ is unknown. Therefore we investigate online estimation techniques in this paper based on observed positions $X_t^\dagger$.

While standard online estimation techniques, such as stochastic gradient descent (SGD) \citep{SS17} and Kalman filtering \citep{nusken2019state,Reich2021}, can be applied in case both ($X_t^\dagger, U_t^\dagger)$ from (\ref{eq:data}) are being observed, the estimation problem becomes more challenging in case only positions $X_t^\dagger$ are being observed and velocities need to be estimated using finite differencing \citep{PLGACF16,BRB20}, such as,
\begin{equation} \label{eq:U approximation}
\tilde U_{n+1/2} := \frac{X_{t_{n+1}}^\dagger -X_{t_n}^\dagger}{\tau}
\end{equation}
with $\tau >0$ denoting the sampling interval, $t_n = \tau n$, for all $n\ge 0$. 

As demonstrated previously \citep{DS04,G06, FCMWG20}, naive application of standard maximum like\-li\-hood-based (MLE) estimators \citep{kutoyants2013statistical,Pavliotis2016} leads to a systematic bias. In order to correct for the arising estimation bias, appropriate corrections terms to the MLE estimator have been proposed in \cite{G06,BRB20,ABGO24, FCMWG20}. An alternative approach has been to treat the continuous time velocity process as a latent variable and calculate the MLE estimator using an expectation-maximization algorithm \cite{BS10}. Viewing the SDEs as a partially observed system, the likelihood of $\theta$ could also be approximated using an extended Kalman filter \cite{Sarkka2013}. Related phenomena have been investigated in the context of multi-scale diffusion \citep{ait2005often,PPS09,BC13}, which again can be corrected via an appropriately modified MLE estimator \citep{DFM2016}. 

While these correction terms can be applied to online learning, we follow an alternative approach in this note, which has been motivated by the related work \citep{AGPSZ21,PRZ23}. 
Our approach can easily be implemented both for SGD \cite{SS17} as well as Kalman filter \cite{nusken2019state,Reich2021} formulations. 

As demonstrated in the next section by means of a simple example, a discretized SGD of the form
\begin{subequations} \label{eq:standard SGD}
\begin{align}
\theta_{n+1} &:= \theta_n + \alpha_t F(X_{t_n}^\dagger,\tilde U_n)^{\rm T}\Delta I_n(\theta_n),
\\
\Delta I_n(\theta) &:= (\tilde U_{n+1/2}-\tilde U_{n-1/2}) - \left( f(X_{t_n}^\dagger,\tilde U_n) + F(X_{t_n}^\dagger,\tilde U_n)\theta \right) \tau,
\end{align}
\end{subequations}
with $U_{t_{n+1/2}}$ being replaced by its estimated $\tilde U_{n+1/2}$ etc.~and the symmetric approximation
\begin{equation} \label{eq:U symmetric}
\tilde U_n := \frac{1}{2}\left(
\tilde U_{n+1/2} + \tilde U_{n-1/2}\right)
= \frac{X_{t_{n+1}}^\dagger-X_{t_{n-1}}^\dagger}{2\tau}
\end{equation}
will fail even in the limit $\tau \to 0$.\footnote{For integers $n, m$, Equation \eqref{eq:U approximation} defines the estimated velocity $\tilde U_{(n + m) + 1/2}$ between time points $n + m$ and $n+m+1$. To keep notation concise, we will write $\tilde U_{n + (2m + 1)/2}$ instead of $\tilde U_{n + m + 1/2}$. This means that $\tilde U_{n - 1/2} = \tilde U_{(n - 1) + 1/2}$, $\tilde U_{n + 3/2} = \tilde U_{(n + 1) + 1/2}$ etc.} Here 
\begin{equation} \label{eq:learning rate}
\alpha_t = c_1/(c_2+t)
\end{equation}
denotes the learning rate for appropriate constants $c_i>0$, $i=1,2$. While the centered approximation in \eqref{eq:U symmetric is the best approximation to $U_n$ from the position measurements at the nearest neighbor time points $X_{t_{n-1}}, X_{t_{n}},X_{t_{n+1}}$ alternative approximations to the velocity at $t_n$ could also be employed.}

The discussion from the following section will lead to an appropriate modification of (\ref{eq:standard SGD}), which removes the undesirable bias in the SGD estimator
(\ref{eq:standard SGD}). See Section \ref{sec:SGD}. 

We extend the proposed methodology to Kalman filter based estimation in Section \ref{sec:Kalman}. The unbiasedness of the corrected SGD approach is demonstrated for a nonlinear second-order diffusion process in Section \ref{sec:Numerics}. The note closes with comments on further directions for research.

%
\section{Motivating example} \label{sec:Example}
%

In order to demonstrate the difficulties the standard SGD implementation (\ref{eq:standard SGD}) encounters, we consider velocity data $U_t^\dagger$ being generated by the linear OU process
\begin{equation} \label{eq:OU}
{\rm d} U_t^\dagger = - \theta_\ast U_t^\dagger {\rm d}t + \sqrt{\sigma}{\rm d}W_t^\dagger
\end{equation}
for given diffusion constant $\sigma>0$ and unknown drift parameter $\theta_\ast \ge 0$. 

We will set $\theta_\ast = 0$ for some of the analysis to be conducted later in this section, which implies that solutions of (\ref{eq:OU}) 
satisfy
\begin{equation}
U_{t}^\dagger = U_s^\dagger + \sqrt{\sigma} W_{t-s}^\dagger 
\end{equation}
for any $t\ge s$ and $s\ge 0$ and, consequently,
\begin{equation}
    X_t^\dagger = X_s^\dagger + (t-s) U_s^\dagger + \sqrt{\sigma}\int_s^t W^\dagger_{r-s}{\rm d}r.
\end{equation}
Here, we have introduced the shorthand $W_{t-s}^\dagger = W_t^\dagger-W_s^\dagger$.

We first recall some of the properties of SGD when a solution of (\ref{eq:OU}) has in fact been fully observed. An application of SGD \citep{SS17} to this problem yields
\begin{subequations} \label{eq:continuous time SGD}
\begin{align}
{\rm d}\theta_t &= -\alpha_t U_t^\dagger {\rm d}I_t =  -\alpha_t U_t^\dagger \left({\rm d}U_t^\dagger + \theta_t U_t^\dagger {\rm d}t \right)\\
&= -\alpha_t U_t^\dagger \left((\theta_t -\theta_\ast) U_t^\dagger {\rm d}t + \sqrt{\sigma}{\rm d}W_t^\dagger \right).
\end{align}
\end{subequations}
We call
\begin{equation}
I_t = \int_0^t \left( {\rm d}U_s^\dagger + \theta_s U_s^\dagger {\rm d}s\right) = \int_0^t \left( 
(\theta_s -\theta_\ast) U_s^\dagger {\rm d}s + \sqrt{\sigma}{\rm d}W_s^\dagger
\right)
\end{equation}
the innovation and $\alpha_t U_t^\dagger$ the gain. It is obvious that $I_t$ becomes a martingale at the optimal parameter choice $\theta_s = \theta_\ast$, $s \in [0,t]$.

It has been shown that
$\theta_t \to \theta_\ast$ as $t\to \infty$ \citep{SS17}. In our context, a key observation is that the integral
\begin{equation}
\int_0^t \alpha_s U_s^\dagger \left({\rm d}U_s^\dagger + \theta_t U_s^\dagger {\rm d}s\right)
\end{equation}
also becomes a martingale for any $t>0$ if and only if $\theta_s = \theta_\ast$ for all $s\in [0,T]$.

In terms of numerical implementations of the SGD formulation (\ref{eq:continuous time SGD}), we consider the discrete time SGD method
\begin{equation}
\theta_{n+1} := \theta_n - \alpha_{t_n} U_{t_n}^\dagger \left( U_{t_{n+1}}^\dagger-U_{t_n}^\dagger +  \frac{\theta_n}{2} \left(U_{t_{n+1}}^\dagger+U_{t_n}^\dagger\right) 
\tau \right).
\end{equation}
This discrete-time SGD method still implies that $\theta_n$ converges to $\theta_\ast$ as $n\to \infty$ and $\tau \to 0$. More precisely, any finite sum
\begin{equation}\label{eq:martingale_trueU}
\sum_{n=0}^{N-1} U_{t_n}^\dagger \left( U_{t_{n+1}}^\dagger-U_{t_n}^\dagger + \frac{\theta_n}{2}\left(U_{t_{n+1}}^\dagger+ U_{t_n}^\dagger \right) \tau \right) 
\end{equation}
becomes a martingale as $\tau \to 0$ for $T=N\tau$ fixed if and only if $\theta_n = \theta_\ast$.

However, the focus of this note is on the case when only positions $X_t^\dagger$, $t\ge 0$, can be observed and we
recall the velocity approximations (\ref{eq:U approximation}), where the sampling interval $\tau>0$ can be made arbitrarily small. We note that (\ref{eq:U approximation}) can also be viewed as a local time average, i.e.,
\begin{equation}
\tilde U_{n+1/2}  = \frac{1}{\tau} \int_{t_n}^{t_{n+1}} U_t^\dagger {\rm d}t,
\end{equation}
which provides a link to the related work \citep{G06,CH20}.

Using (\ref{eq:U symmetric}) instead of the true $U_{t_n}^\dagger$ etc.~ and symmetrizing the gain leads to the previously stated (\ref{eq:standard SGD}), which reduces here to
\begin{subequations} \label{eq:SGD OU}
\begin{align}
\theta_{n+1} &:= \theta_n - \alpha_{t_n} \tilde U_n \left( \tilde U_{n+1/2}-\tilde U_{n-1/2} + \theta_n \tilde U_{n}  \tau \right)\\
&= \theta_n - \frac{\alpha_{t_n}}{2} \left(X_{t_{n+1}}^\dagger-X_{t_{n-1}}^\dagger\right) \left( \frac{X_{t_{n+1}}^\dagger-2X_{t_{n}}^\dagger+X_{t_{n-1}}^\dagger }{\tau^2}
+ \frac{\theta_n}{2} \frac{X_{t_{n+1}}^\dagger-X_{t_{n-1}}^\dagger}{\tau} \right).
\end{align}
\end{subequations}
We note that, while the discrete innovations
\begin{equation}
I_N = \sum_{n=1}^{N-1} \left( \frac{X_{t_{n+1}}^\dagger-2X_{t_n}^\dagger+X_{t_{n-1}}^\dagger}{\tau^2}
+ \frac{\theta_n}{2} \frac{X_{t_{n+1}}^\dagger-X_{t_{n-1}}^\dagger}{\tau} \right)\tau
\end{equation}
still converge to a martingale as $\tau \to 0$ for $T = N\tau$ fixed provided $\theta_n = \theta_\ast$ for all $n\ge 0$, this is no longer true for 
\begin{equation} \label{eq:sum}
\frac{1}{2}\sum_{n=1}^{N-1} \left(X_{t_{n+1}}^\dagger-X_{t_{n-1}}^\dagger\right) \left( \frac{X_{t_{n+1}}^\dagger-2X_{t_n}^\dagger+X_{t_{n-1}}^\dagger }{\tau^2}
+ \frac{\theta_n}{2} \frac{X_{t_{n+1}}^\dagger-X_{t_{n-1}}^\dagger}{\tau} \right).
\end{equation}
This violation of the martingale property is perhaps not surprising since (\ref{eq:sum}) does not resemble a discretized It\^o-Integral like (\ref{eq:martingale_trueU}) but rather a discretized Stratonovitch integral \citep{Pavliotis2016}. We note that such an interpretation as a discretized stochastic integral concerns only the second-order differencing term in (\ref{eq:sum}) and we may therefore set $\theta=\theta_\ast = 0$ in the subsequent derivations. Recall that
\begin{equation}
\frac{X_{t_{n+1}}-X_{t_n}}{\tau} = \tilde U_{n+1/2} = U_{t_{n}}^\dagger + \frac{\sqrt{\sigma}}{\tau} \int_{t_n}^{t_{n+1}} W_{t-t_{n}}^\dagger {\rm d}t 
\end{equation}
and we therefore obtain
\begin{subequations}
\begin{align}
\frac{1}{2} \mathbb{E}\left[ \left(\tilde U_{n+1/2}+\tilde U_{n-1/2} \right) \left(\tilde U_{n+1/2}-\tilde U_{n-1/2}\right) \right]
&= \frac{1}{2} \mathbb{E}\left[\tilde U_{n+1/2}^2  -  \tilde U_{n-1/2}^2 \right] \\
&= \frac{1}{2} 
\mathbb{E} \left[ (U_{t_n}^\dagger)^2 -(U_{t_{n-1}}^\dagger)^2 \right] = \frac{\tau \sigma}{2},
\end{align}
\end{subequations}
which implies an estimation bias induced by the discrete sum (\ref{eq:sum}) even in the limit $\tau \to 0$.

In order to remove the resulting bias in the SGD formulation (\ref{eq:SGD OU}), we propose to simply shift the innovation term forward in time such that it does no longer overlap with the gain; thus regaining an It\^o-type stochastic integral approximation. A related idea has been proposed and investigated in \cite{PRZ23} in the context of multi-scale diffusion. This modification leads to the following implementation:
\begin{equation} \label{eq:SGD}
\theta_{n+1} := \theta_n + \alpha_{t_n} \tilde U_{n} \left( \tilde U_{n+5/2}-\tilde U_{n+3/2} - 
\theta_n \tilde U_{n+2}\tau \right).
\end{equation}
Note that the innovation in the modified implementation (\ref{eq:SGD}) only contains positional data from time points that are larger than or equal to those in the gain.

Let us introduce the random variables
\begin{equation}
\tilde \Xi_{n} := \frac{1}{\sqrt{\sigma \tau}} \left( \tilde U_{n+1/2} - \tilde U_{n-1/2} \right)
\end{equation}
for $n\ge 1$, which is a rescaling of the innovation $\Delta I_n(\theta_\ast)$, as defined in (\ref{eq:standard SGD}b), with
$\theta_\ast$ again set to $\theta_\ast=0$.  It is crucial to observe that $\tilde \Xi_{n+2}$ is correlated with both $\tilde U_{n+5/2}$ and $\tilde U_{n+3/2}$, while being independent of $\tilde U_{n}$. Hence, provided $\tilde \Xi_{n}$ is a martingale for all $n \ge 1$, the SGD scheme (\ref{eq:SGD}) will become unbiased in the limit $\tau \to 0$.

Hence let us analyze the statistical properties of $\tilde \Xi_n$.  We find that
\begin{subequations} \label{eq:Xi}
\begin{align}
\tilde \Xi_n &= \frac{1}{\sqrt{\sigma \tau}} \left( \tilde U_{n+1/2} - \tilde U_{n-1/2} \right) \\ &=
\frac{1}{\sqrt{\sigma \tau}} \left( U_{t_n}^\dagger- U_{t_{n-1}}^\dagger\right) + \frac{1}{\sqrt{\tau}} \left(
\frac{\int_{t_n}^{t_{n+1}} W_{t-t_n}^\dagger {\rm d}t - \int_{t_{n-1}}^{t_n} W_{t-t_{n-1}}^\dagger {\rm d}t}{\tau}\right)\\
&= \frac{W_{t_n-t_{n-1}}^\dagger}{\sqrt{\tau}} + \frac{1}{\sqrt{\tau}} \left(
\frac{\int_{t_n}^{t_{n+1}} W_{t-t_n}^\dagger {\rm d}t - \int_{t_{n-1}}^{t_n} W_{t-t_{n-1}}^\dagger {\rm d}t}{\tau}\right).
\end{align}
\end{subequations}
It follows that $\tilde \Xi_n$ is Gaussian with mean zero and that $\tilde \Xi_{n+2}$ is indeed independent of $\tilde U_n$.

We recall the following two properties of integrated Brownian motion:
\begin{equation} 
\mbox{var}\,\left(\int_{t_n}^{t_{n+1}} W_{t-t_n}^\dagger {\rm d}t\right) = \frac{1}{3}\tau^3
\end{equation}
and
\begin{equation}
\mathbb{E}\left[ W_{t_n-t_{n-1}}^\dagger\left(\int_{t_{n-1}}^{t_n} W_{t-t_{n-1}}^\dagger {\rm d}t\right)\right] = \int_{t_{n-1}}^{t_n}
\mathbb{E} \left[ (W_{t-t_{n-1}}^\dagger)^2 \right] {\rm d}t
= \frac{1}{2}\tau^2.
\end{equation}

Upon applying these results to (\ref{eq:Xi}c) when computing the variance of $\tilde \Xi_n$, it is found that
\begin{equation}
\mbox{var}\,(\tilde \Xi_n) = \frac{2}{3}. \label{eq:Xi_variance}
\end{equation}
In other words, we obtain that $\tilde \Xi_n \sim {\rm N}(0,2/3)$ in agreement with results from \cite{G06,CKP24}. This result carries over to $\theta_\ast \not= 0$ since the arising additional terms vanish as $\tau \to 0$. 

We finally formally investigate the time-continuous limit of the modified discrete-time SGD formulation (\ref{eq:SGD}). We first note that
\begin{equation}
\mathbb{E}[\tilde \Xi_n \tilde \Xi_{n+k}]
= \left\{ \begin{array}{cc} \frac{1}{6} & |k|= 1\\
0 & |k| > 1
\end{array} \right. ,
\end{equation}
which implies that
\begin{equation}
Z_t \mathrel{\mathop:}=
\begin{cases}
0 & \text{if} \quad t < \tau\\
     \sum_{k=1}^{\lfloor t/ \tau \rfloor}\sqrt{\tau}\,\tilde{\Xi}_k & \text{else}
\end{cases}
\end{equation}
converges to standard Brownian motion in the limit $\tau \to 0$. When using this result in the SGD formulation (\ref{eq:SGD}) for general $\theta_\ast$, we obtain
\begin{subequations}
\begin{align}
\theta_{n+1} &\approx \theta_n - \alpha_{t_n} \tilde U_{n} \left( 
(\theta_n-\theta_\ast)\tilde U_{n+2} \tau + \sqrt{\sigma \tau} \,\tilde \Xi_{n+2} \right) \\
&\approx  \theta_n - \alpha_{t_n} \tilde U_{n} \left( 
(\theta_n-\theta_\ast) \tilde U_n   \tau  
+  \sqrt{\sigma \tau}\, \tilde \Xi_{n+2} \right)
\end{align}
\end{subequations}
and the limit $\tau \to 0$ leads formally to the time-continuous formulation
\begin{equation}
\dot{\theta}_t = -\alpha_t U_t^\dagger \left((\theta-\theta_\ast) U_t^\dagger + \sqrt{\sigma} \dot{W}_t^\dagger\right),
\end{equation}
which coincides with (\ref{eq:continuous time SGD}b).

\textit{Remark: Independent of the drift estimation, it is well known that the diffusion term can be estimated from the variability of a process.  In line with previous results \cite{CKP24, FCMWG20, G06}, our above calculations and equation \eqref{eq:Xi_variance} imply the following estimator for the diffusion constant $\sigma$ from the quadratic variation of the approximated velocities $\tilde{U}_{n+1/2}$:
\begin{equation} \label{eq:sigma estimate}
\sigma_N := \frac{3}{2\tau N} \sum_{n=1}^N \|\tilde U_{n+1/2} - \tilde U_{n-1/2}\|^2.
\end{equation}}

%
\section{Unbiased SGD} \label{sec:SGD}
%

The analysis from Section \ref{sec:Example} leads to the following modification of the discrete-time SGD implementation
(\ref{eq:standard SGD}):
\begin{equation} \label{eq:new SGD}
\theta_{n+1} = \theta_n + \alpha_{t_n} F(X_{t_n}^\dagger,\tilde U_n)^{\rm T}\Delta I_{n+2}(\theta_n),
\end{equation}
for $n\ge 1$ with the innovations $\Delta I_n(\theta)$ defined by (\ref{eq:standard SGD}b) and the learning rate $\alpha_{t_n}$ by (\ref{eq:learning rate}) with $t = t_n = n\tau$. We recall that the key point of shifting the innovation forward in time is to retain an It\^o-type approximation in the gain times innovation stochastic integral approximation. This is the property that maintains the desired unbiasedness of the resulting sequential learning scheme in the limit $\tau \to 0$. 

We note that unbiased formulations are not unique since only the shifting of the second-order differecing term in the innovation is relevant for consistency. For example, (\ref{eq:new SGD}) could be replaced by
\begin{equation}
    \theta_{n+1} = \theta_n + \alpha_{t_n}
    F(X_{t_n}^\dagger,\tilde U_n)^{\rm T} \left(
    \tilde U_{n+5/2}- \tilde U_{n+3/2}
    - \left( f(X_{t_n}^\dagger,\tilde U_n) +
    F(X_{t_n}^\dagger,\tilde U_n) \theta_n \right) \tau \right).
\end{equation}
This freedom is also utilized when formulating the modified MLE estimator (\ref{eq:new MLE}) further below in Section \ref{sec:conclusion}.

%
\section{Unbiased Kalman filter} \label{sec:Kalman}
%

The SGD method (\ref{eq:new SGD}) provides an online point estimator for the parameter $\theta$. However, it is also possible to implement an online Bayesian estimator using the Kalman filter \citep{nusken2019state,Reich2021}. The key idea is to treat the unknown parameter as a Gaussian random variable $\Theta_n$ with mean $m_n$ and covariance matrix $\Sigma_n$ and with trivial dynamics $\Theta_{n+1} = \Theta_{n}$, $n\ge 0$ \cite{Sarkka2013}. The evolution equation for the mean and the covariance matrix follow the standard Kalman update formulas:
\begin{subequations} \label{eq:Kalman}
    \begin{align}
    m_{n+1} &:= m_n + K_n
    \frac{\Delta I_{n+2}(m_n)}{\tau},\\
    \Sigma_{n+1} &:= \Sigma_n - K_n
     F(X_{t_n}^\dagger,\tilde U_n)
    \Sigma_n
    \end{align}
\end{subequations}
with Kalman gain
\begin{equation} \label{eq:gain}
K_n := 
    \Sigma_n F(X_{t_n}^\dagger,\tilde U_n)^{\rm T}
    \left(F(X_{t_n}^\dagger,\tilde U_n)^{\rm T} \Sigma_n
    F(X_{t_n}^\dagger,\tilde U_n) + \frac{\sigma}{\tau}I\right)^{-1}
\end{equation}
for $n\ge 1$. The iteration is initialized from the Gaussian prior $\Theta_1 \sim {\rm N}(m_{\rm 
prior},\Sigma_{\rm prior})$.

In contrast to the SGD formulation (\ref{eq:new SGD}), the learning rate $\alpha_{t_n}$ no longer appears and has instead been submerged implicitly into the Kalman gain $K_n$. Indeed, the singular values of the covariance matrix $\Sigma_n$ decay with rate $1/t_n$ under the already stated ergodicity assumption on (\ref{eq:data}) and the non-singularity of the Fisher information matrix $\mathbb{E}_\pi\left[F(X,U)^{\rm T}F(X,U)\right]$. If the required diffusion constant $\sigma$ is unknown, it can either be estimated via (\ref{eq:sigma estimate}) or, alternatively, be treated as part of the prior covariance matrix $\Sigma_{\rm prior}$ and then set to $\sigma = 1$ in the Kalman gain (\ref{eq:gain}).

%
\section{Numerical example} \label{sec:Numerics}
%

\begin{figure}[htbp] 
\includegraphics[width = \textwidth]{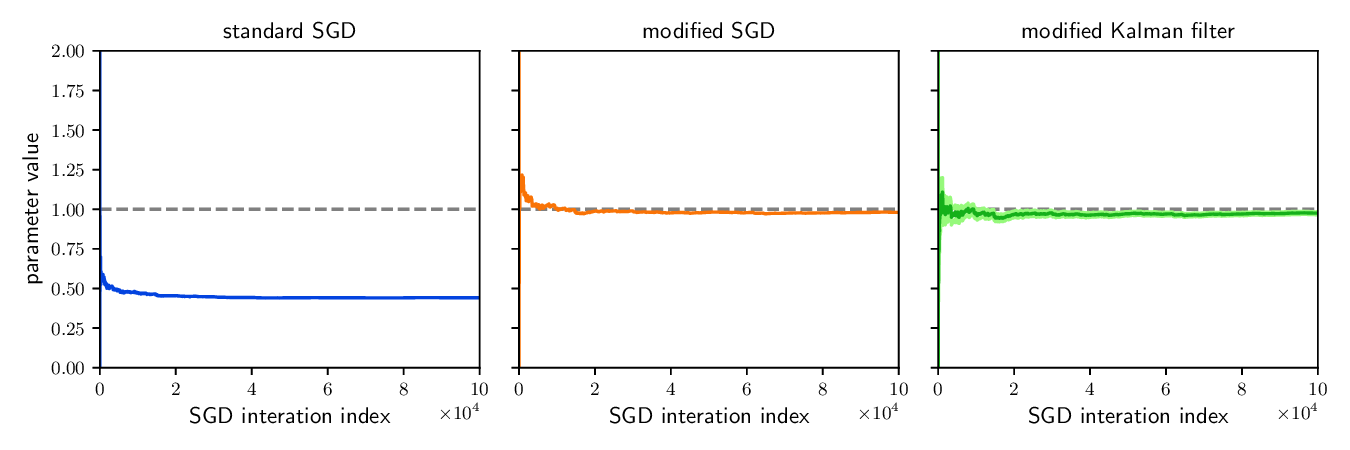}
\caption{Left panel: Estimated parameter value from standard SGD as a function of the iteration index. The iteration starts at $\theta_0 = 2$ and the reference value is $\theta_\ast = 1$. A systematic bias can clearly be seen. Middle panel: Same data is used in the modified SGD. The modified SGD algorithm converges to the correct reference value. Right panel: Same data is used in the unbiased Kalman filter with prior distribution $N(2, 1)$ and $\sigma$ treated as part of the prior. The lightly colored lines indicate the $1\sigma$-intervals of the Bayesian estimates.}
\label{fig1}
\end{figure}

We consider the nonlinear second-order SDE
\begin{subequations}
    \begin{align}
        {\rm d}U_t &= -X_t{\rm d}t +
        U_t {\rm d}t - \theta U_t^3 {\rm d}t
        + \sqrt{2}{\rm d}W_t,\\
        {\rm d}X_t &= U_t {\rm d}t
    \end{align}
\end{subequations}
with unknown parameter $\theta \in \mathbb{R}$.
The data $X_t^\dagger$ is generated with $\theta_\ast = 1$, while the SGD iteration starts from $\theta_0= 2$. The sample interval is set to $\tau = 0.025$. The data is generated with a much smaller step-size of $h = \tau/100$. A total of $N = 10^5$ positions $X_{t_n}$, $n=1,\ldots,N$ have been generated. The learning rate is given by $\alpha_{t_n} = 6.0/n$. See Figure \ref{fig1} middle panel for the time evolved estimates $\theta_n$ generated by the modified SGD (\ref{eq:new SGD}). These estimates clearly converge to $\theta_\ast$. In Figure \ref{fig1} we also display results from the standard SGD as well as the modified Kalman filter presented in section \ref{sec:Kalman}. For the Kalman filter, a Gaussian prior was used with $m_{\rm{prior}}=2$ and $\Sigma_{\rm{prior}} = 6$. The diffusion constant $\sigma$ was assumed to be unknown and treated as part of the prior. While the standard SGD converges slightly faster than the modified SGD, it converges to a wrong parameter value. The estimates of the modified Kalman filter are very similar to estimates of the modified SGD. An unmodified Kalman filter (not shown) on the other hand leads to a similar estimation bias as the standard SGD.

%
\section{Conclusions} \label{sec:conclusion}
%

In this note, we have proposed a simple-to-implement modification of standard SGD and Kalman filter formulations for online parameter estimation from position data only. The essential idea is to establish an It\^o-like modification that re-establishes the martingale property in the presence of temporal correlations. The approach has been inspired by related work \citep{AGPSZ21,PRZ23} on multi-scale diffusion and parameter estimation for the arising reduced equations. We finally note that a corrected MLE estimator can be based on the negative log-likelihood function
\begin{subequations} \label{eq:new MLE}
    \begin{align}
    \mathcal{L}(\theta,\sigma) &=
    \frac{1}{2\sigma}
    \sum_{n=1}^{N-3} \theta^{\rm T} F(X_{t_n}^\dagger,\tilde U_n)^{\rm T}F(X_{t_{n}}^\dagger,\tilde U_{n})\theta \tau \\
    & \qquad -\,\frac{1}{\sigma} \sum_{n=1}^{N-3} \theta^{\rm T} F(X_{t_n}^\dagger,\tilde U_n)^{\rm T} \left( \tilde U_{n+5/2}-\tilde U_{n+3/2} - f(X_{t_n}^\dagger,\tilde U_{n})\tau\right)\\
    & \qquad +\,\frac{3}{4\tau \sigma} \sum_{n=1}^{N-3}
    \| \tilde U_{n+5/2}- \tilde U_{n+3/2}\|^2 +
    \frac{N-3}{2}\log \sigma .
    \end{align}
\end{subequations}
Compare this likelihood function to the contrast function presented in \cite{G06}, which is based on adding an appropriate correction term. Please also note that (\ref{eq:new MLE}a)-(\ref{eq:new MLE}b) replaces the ergodic least squares functional (\ref{eq:least squares}) using measured positions and estimated velocities and that (\ref{eq:new MLE}c) delivers the estimator (\ref{eq:sigma estimate}).

Future directions include the investigation of the proposed estimators in the context multi-scale diffusion when (\ref{eq:SDE}) arises from homogenization of an underlying multi-scale diffusion process and the inclusion of observation errors in the positions $X_t$. In the former case, a combination of the approaches taken here and in \cite{PRZ23} emerges as a feasible starting point. 

\medskip

\paragraph{Acknowledgments.}

This work has been partially funded by Deutsche Forschungsgemeinschaft (DFG) - Project-ID 318763901 - SFB1294.  

\bibliographystyle{plainurl}
%
\bibliography{refs_estimation}
%
 
\end{document}